\documentclass[aps,prl,twocolumn,groupedaddress]{revtex4}
\usepackage{graphicx}
\begin{document}

\title{Heat capacity at the glass transition}
\author{Kostya Trachenko$^{1,2}$}
\author{V. V. Brazhkin$^{3}$}
\address{$^1$ Department of Physics, Queen Mary University of London, Mile End Road, London, E1 4NS, UK}
\address{$^2$ Department of Earth Sciences, University of Cambridge, Cambridge CB2 3EQ, UK}
\address{$^3$ Institute for High Pressure Physics, RAS, 142190, Troitsk, Moscow Region, Russia}

\begin{abstract}
A fundamental problem of glass transition is to explain the jump of heat capacity at the glass transition temperature $T_g$
without asserting the existence of a distinct solid glass phase. Similar problems are also common to other disordered systems,
including spin glasses. We propose that if $T_g$ is defined as the temperature at which the liquid stops relaxing at the
experimental time scale, the jump of heat capacity at $T_g$ follows as a necessary consequence due to the change of the liquid's
elastic, vibrational and thermal properties. In this picture, we discuss time-dependent effects of glass transition, and
identify three distinct regimes of relaxation. Our approach explains widely observed logarithmic increase of $T_g$ with the
quench rate and correlation of the heat capacity jump with liquid fragility.
\end{abstract}


\maketitle

\section{Introduction}

When a transition takes place between two distinct phases, the change of heat capacity and other thermodynamic quantities is
consistently understood in a theory of phase transitions \cite{landau}. Often a disordered system such as a liquid forms a
similarly disordered solid glass without a transition into a different phase, yet heat capacity changes with a jump. The jump is considered a hallmark of glass transition, and defines glass transition temperature $T_g$. The heat capacity jump immediately presents a problem that is at the heart of glass transition \cite{dyre,angell}: how can the jump be understood if there is no distinct second phase?

This problem remains unsolved and controversial. One set of theories rationalize the jump of heat capacity by invoking thermodynamics of phase transitions. An instructive illustration is the ongoing discussion of a popular theory of glass transition, the Adam-Gibbs theory \cite{adam}. The theory connects the change of heat capacity at $T_g$ to the configurational entropy which becomes zero below $T_g$ where a phase transition between a liquid and a glass takes place \cite{dyre,dyre1}. The Adam-Gibbs theory has been convincingly criticized for a number of important reasons \cite{dyre1}. Chief of these, also present in other similar theories, is that it has not been possible to identify the second low-temperature phase (the glass phase). To circumvent this problem, several theories have subsequently put forward the proposals about the non-conventional mechanisms of the phase transition and non-trivial descriptions of the second phase, while retaining the idea of a phase transition of sort \cite{dyre}.

Another set of glass transition theories consider that glass transition phenomena at $T_g$ have purely dynamic origin, and simply correspond to the freezing of atomic jumps in a liquid at the experimental time scale \cite{dyre}. The absence of a phase transition and thermodynamic effects at $T_g$ is supported by the wide experimental observation that the liquid and the glass at $T_g$ have nearly identical structure \cite{dyre}. However, the challenge for the dynamic theories is to explain both the origin of the jump of heat capacity at $T_g$ and its large magnitude, which for some systems can be of the order of $k_{\rm B}$ per atom.

In addition to glass transition in structural liquids, similar problems exist in other disordered media. For example, spin
glasses have seen large developments of ideas based on phase transitions and the existence of the second distinct phase.
Similar to the structural glass transition, these theories have been used to explain the cusp in susceptibility at the glass
transition temperature. Similar to the structural glass transition, several important problems remain in this field as well,
including identifying the nature of a distinct spin glass phase, dependence of the cusp on field frequency or observation
time, slow relaxation effects etc \cite{mydosh}.

In this paper, we propose how to explain the jump of heat capacity in a purely dynamic picture, without asserting the existence of a distinct thermodynamic solid glass phase and, therefore, show how to reconcile the above contradiction. We recall that glass transition temperature $T_g$ has two experimental definitions which give similar values of $T_g$. In the calorimetry experiments, $T_g$ is the temperature at which the jump of constant-pressure heat capacity, $C_p$, is seen. In the experiments that measure $\tau$ (e.g. dielectric relaxation experiments), $T_g$ is the temperature at which $\tau$ exceeds the time of the experiment $t$ of about $10^2-10^3$ s. We propose that when $\tau$ exceeds $t$, the jump of heat capacity at $T_g$ follows as a necessary consequence because freezing of local relaxation events alters liquid elastic, vibrational and thermal properties including bulk modulus and thermal expansion. Hence, there is no need to invoke the existence of a second glass phase and a phase
transition of sort. In this picture, we discuss time-dependent effects of glass transition, and identify three distinct
regimes of relaxation. Our approach explains widely observed logarithmic increase of $T_g$ with the quench rate and the
correlation of heat capacity jump with liquid fragility.

\section{Change of heat capacity at $T_g$}

The commonly discussed quantity from the calorimetry experiments is the ratio of constant-pressure liquid heat capacity,
$C_p^l$, to glass heat capacity, $C_p^g$. We do not consider the overshoot of heat capacity on heating and its undershoot on cooling, discussed elsewhere \cite{moyn1,sethna}. Heat capacities are considered at temperatures separated by the interval in which these effects decay to the values of $C_p^l$ and $C_p^g$ attributed to the liquid and the glass \cite{angell,wang1}. For various liquids, $\frac{C_p^l}{C_p^g}=1.1-1.8$ \cite{angell}. Using the known relationship $C_p-C_v=VT\alpha^2B$, where $C_v$ is the constant-volume heat capacity, $\alpha$ is the coefficient of thermal expansion and $B$ is bulk modulus, we write

\begin{equation}
\frac{C_p^l}{C_p^g}=\frac{C_v^l+V_lT_l\alpha_l^2B_l}{C_v^g+V_gT_g\alpha_g^2B_g} \label{1}
\end{equation}

\noindent where $_l$ and $_g$ subscripts refer to the liquid and the glass.

We start with addressing the origin of the difference between $B_l$ and $B_g$ and between $\alpha_l$ and $\alpha_g$. Unlike
in a solid glass, atoms in a liquid are not fixed, but rearrange in space. This gives liquid flow. Each flow event is a jump
of an atom from its surrounding cage, accompanied by large-scale rearrangement of the cage atoms. We call this process a
local relaxation event (LRE). A LRE lasts on the order of Debye vibration period $\tau_0=0.1$ ps. Frenkel introduced liquid
relaxation time $\tau$ as the time between LREs at one point in space in a liquid \cite{frenkel}. Frenkel's main idea was
that at short times $t<\tau$, liquid response is the same as that of a solid, i.e. is purely elastic. On the other hand, for
$t>\tau$, viscous flow takes place, during which each LRE is accompanied by additional, viscous, displacement. Hence, for
$t>\tau$, liquid response to external perturbation (e.g. pressure) consists of elastic and viscous response \cite{frenkel}.
This discussion provided the microscopic basis for the earlier phenomenological model by Maxwell \cite{maxw}, who proposed
to separate elastic and viscous response in his viscoelastic approach to liquid flow.

Lets consider that pressure $P$ is applied to a liquid. Pressure induces a certain finite number of LREs, which bring the
liquid to the equilibrium state at new external conditions ($P$, $T$) after time $\tau$. Following the Maxwell-Frenkel
approach, the change of liquid volume, $v$, is $v=v_{el}+v_{r}$, where $v_{el}$ and $v_r$ are associated with solid-like
elastic deformation and viscous relaxation process due to LREs, respectively. Lets now define $T_g$ as the temperature at
which $\tau$ exceeds the observation time $t$. This implies that LREs are not operative at $T_g$ during the time of
observation. Therefore, $v$ at $T_g$ is given by purely elastic response as in elastic solid. Then, $P=B_l\frac{v_{el}+v_r}{V_l^0}$ and $P=B_g\frac{v_g}{V_g^0}$ where $V_l^0$ and $V_g^0$
are initial volumes of the liquid and the glass and $v_g$ is the elastic deformation of the glass. Let $\Delta T$ be a small
temperature interval that separates the liquid from the glass such that $\tau$ in the liquid, $\tau_l$, is
$\tau_l=\tau(T_g+\Delta T)$ and $\frac{\Delta T}{T_g}\ll 1$. Then, $V_l^0\approx V_g^0$. Similarly, the difference between
the elastic response of the liquid and the glass can be ignored for small $\Delta T$, giving $v_{el}\approx v_g$.
Combining the two expressions for $B_l$ and $B_g$, we find:

\begin{equation}
B_l=\frac{B_g}{\epsilon_1+1} \label{2}
\end{equation}

\noindent where $\epsilon_1=\frac{v_{r}}{v_{el}}$ is the ratio of relaxational and elastic response to pressure.

$\alpha_l$ can be calculated in a similar way. Lets consider liquid relaxation in response to the increase of temperature by
$\Delta T$. We write $\alpha_l=\frac{1}{V_0^l}\frac{v_{el}+v_r}{\Delta T}$ and $\alpha_g=\frac{1}{V_0^g}\frac{v_g}{\Delta T}$, where $v_{el}$ and $v_r$ are temperature-induced volume increases in a liquid that are related to solid-like elastic and relaxational response, respectively, and $v_g$ is the elastic response of the glass. Combining the two expressions and assuming $V_l^0=V_g^0$ and $v_{el}=v_g$ as before, we find

\begin{equation}
\alpha_l=(\epsilon_2+1)\alpha_g \label{3}
\end{equation}

\noindent where $\epsilon_2=\frac{v_r}{v_{el}}$ is the ratio of relaxational and elastic response to temperature variation.

Eqs. (\ref{2},\ref{3}) describe the relationships between $B$ and $\alpha$ in the liquid and the glass due to the presence of LREs
in the liquid above $T_g$ and their absence in the glass at $T_g$, insofar as $T_g$ is the temperature at which $t<\tau$.
We note that relaxational response $v_r$ of both liquid and glass decays during time $\tau$. Because $B_g$ and $\alpha_g$ correspond to $t<\tau$,
$B_g$ and $\alpha_g$ are unrelaxed, or non-equilibrium values of bulk modulus and thermal expansion, respectively. This point is discussed in
Chapter 3 in detail.

Using Eqs. (\ref{2},{3}) in Eq. (\ref{1}), we write:

\begin{equation}
\frac{C_p^l}{C_p^g}=\frac{\frac{C_v^l}{C_v^g}+\gamma\alpha_g T_g\epsilon}{1+\gamma\alpha_g T_g} \label{4b}
\end{equation}

\noindent where $\epsilon=\frac{(\epsilon_2+1)^2}{\epsilon_1+1}$, $\gamma=V_g\alpha_g B_g/C_v^g$ is the glass
Gr\"{u}neisen parameter and $C_p^g$ is the unrelaxed, or non-equilibrium (see Chapter 3) heat capacity of the glass.
We note that in Eq. (\ref{4b}), as in Eqs. (\ref{2},\ref{3}), we set $T_l\approx T_g$ and
$V_l\approx V_g$. This underestimates the experimental $\frac{C_p^l}{C_p^g}$, setting its lower limit, because $C_p$ is
measured in the finite range of temperature and volume such that $T_l>T_g$ and $V_l>V_g$.

We now calculate $\frac{C_v^l}{C_v^g}$. Close to $T_g$, $C_v^l$ is due to the vibrational motion only, whereas the
contribution to $C_v^l$ due to the diffusional motion is negligible. This is an important assertion that perhaps was not
appreciated before, and simplifies the problem greatly. The assertion follows from the explicit calculation of liquid $C_v$
as a function of $\tau$ \cite{trac}, and is consistent with the experimental results showing that liquid $C_v$ is close
to $3N$ around the melting point \cite{grimvall,wallace}. It also follows from a more general argument that does not rely on
the explicit calculation of $C_v^l$, as discussed below.

Above $T_g$, each atom participates in the vibrational motion during time $\tau$ and in the diffusional motion when it jumps
between two equilibrium positions during time of about $\tau_0$. For $t\gg\tau$, atoms can be separated into two ensembles
of $N_{vib}$ vibrating and $N_{dif}$ diffusing atoms. The energy of an atom in each ensemble consists of kinetic and
potential energy. Then, the partition sum is $Z=Z_{vib}\cdot Z_{dif}$, where $Z_{vib}$ is due to vibrations and $Z_{dif}$ is
due to diffusion. Liquid energy is $E=T^2\frac{\rm d}{{\rm d}T}\left(\ln(Z_{vib}\cdot Z_{dif})\right)=T^2\frac{\rm d}{{\rm
d}T}\ln Z_{vib}+T^2\frac{\rm d}{{\rm d}T}\ln Z_{dif}=E_{vib}+E_{dif}$, where $E_{vib}$ and $E_{dif}$ is the energy of
vibrating and diffusing atoms, respectively. Here and below, temperature derivatives are taken at constant volume. At any
given moment of time, $N_{dif}=N_0\exp(-U/T)$, where $N_0$ is the total number of atoms in a system and $U$ is the
activation energy barrier for a LRE ($U$ can be temperature-dependent). Combining it with $\tau=\tau_0\exp(U/T)$, where
$\tau_0$ is the Debye vibrational period of about 0.1 ps, we write

\begin{equation}
N_{dif}=N_0\frac{\tau_0}{\tau}
\end{equation}

\noindent which also directly follows by noting that the jump probability is $\tau_0/\tau$.

At $T_g$, $\frac{\tau_0}{\tau}\approx 10^{-16}$, i.e. the number of diffusing atoms is negligible. Therefore, the ratio of
the energy due to diffusion to the total energy, $\frac{E_{dif}}{E}$, is negligible. Similarly, $\frac{E_{dif}}{E_{vib}}\ll
1$, giving

\begin{equation}
\frac{\frac{\rm d}{{\rm d}T}\ln Z_{dif}}{\frac{\rm d}{{\rm d}T}\ln Z_{vib}}\ll 1 \label{4c}
\end{equation}

Liquid entropy is $S=\frac{\rm d}{{\rm d}T}\left(T\ln(Z_{vib}\cdot Z_{dif})\right)=\left(T\frac{\rm d}{{\rm d}T}\ln
Z_{vib}+\ln Z_{vib}\right)+\left(T\frac{\rm d}{{\rm d}T}\ln Z_{dif}+\ln Z_{dif}\right)=T\frac{\rm d}{{\rm d}T}\ln
Z_{vib}+\ln Z_{vib}+\ln Z_{dif}$, where we have used Eq. (\ref{4c}). Then, $C_v^l=T\frac{{\rm d}S}{{\rm d}T}=T\frac{\rm
d}{{\rm d}T}\left(T\frac{\rm d}{{\rm d}T}\ln Z_{vib}\right)+T\frac{\rm d}{{\rm d}T}\ln Z_{vib}+T\frac{\rm d}{{\rm d}T}\ln
Z_{dif}=T\frac{\rm d}{{\rm d}T}\left(T\frac{\rm d}{{\rm d}T}\ln Z_{vib}\right)+T\frac{\rm d}{{\rm d}T}\ln Z_{vib}$, where we
have used Eq. (\ref{4c}) once more. Therefore, $C_v^l$ around $T_g$ (as well as at any temperature $T$ such that
$\frac{\tau_0}{\tau(T)}\ll 1$) is essentially due to the vibrational contribution to $Z$.

The vibrational states of a liquid are given by one longitudinal mode and two transverse modes with frequency
$\omega>1/\tau$ \cite{frenkel}. If $\frac{\tau_0}{\tau}\ll 1$, as is the case around $T_g$, transverse modes in a liquid
account for essentially all transverse modes that exist in a solid glass. Together with the fact that the phonon density of
states increases as $\propto\omega^2$, this means that the energy of the missing transverse waves with frequency
$\omega<1/\tau$ is negligible. Hence, the vibrational energy of a liquid in the regime $\frac{\tau_0}{\tau}\ll 1$ can be
calculated as the energy of all $3N$ phonons as in a solid glass \cite{trac}. Therefore, in discussing the vibrational $C_v$
of a liquid around $T_g$, we can use the results derived for solids.

The partition function of a harmonic solid is $Z=\left(\frac{T}{\hbar\omega}\right)^{3N}$ ($k_{\rm B}$=1), where $\omega$ is the
geometrically averaged phonon frequency \cite{landau}, giving the free energy $F=3NT\ln\frac{\hbar\omega}{T}$. In the purely
harmonic case, $\omega$ is constant, giving the entropy $S=-\left(\frac{\partial F}{\partial
T}\right)_v=3N\left(1+\ln\frac{T}{\hbar\omega}\right)$ and $C_v=T\left(\frac{\partial S}{\partial T}\right)_v=3N$. On the
other hand, anharmonicity, particularly large in liquids, results in the decrease of $\omega$ with temperature. Importantly,
as we show below, this decrease is different below and above $T_g$ because $\alpha$ is different (see Eq. (\ref{3})). If
$\omega$ is not constant, $S=3N\left(1+\ln\frac{T}{\hbar\omega}-\frac{T}{\omega}\frac{{\rm d}\omega}{{\rm d}T}\right)$, and

\begin{equation}
C_v=3N\left(1-\frac{2T}{\omega}\frac{{\rm d}\omega}{{\rm d}T}+ \frac{T^2}{\omega^2}\left(\frac{{\rm d}\omega}{{\rm
d}T}\right)^2-\frac{T^2}{\omega}\frac{{\rm d^2}\omega}{{\rm d}T^2}\right) \label{4d}
\end{equation}

\noindent where the derivatives are taken at constant volume.

The effect of anharmonicity can be discussed in the quasi-harmonic approximation by introducing the Gr\"{u}neisen parameter
$\gamma=-\frac{V}{\omega}\left(\frac{\partial\omega}{\partial V}\right)_T$ to the phonon pressure,
$P_{ph}=-\left(\frac{\partial F}{\partial V}\right)_T=\frac{3NT\gamma}{V}$. Then, the bulk modulus due to the (negative)
phonon pressure is $B_{ph}=-\frac{3NT\gamma}{V}$ and $\left(\frac{\partial B_{ph}}{\partial
T}\right)_v=-\frac{3N\gamma}{V}$, where we neglected the dependence of $\gamma$ on $V$. Using $\gamma=\frac{V\alpha B}{C_v}$
and $B=B_0+B_{ph}$, where $B_0$ is the zero-temperature bulk modulus, $\left(\frac{\partial B_{ph}}{\partial
T}\right)_v=-\alpha(B_0+B_{ph})$, where we set $C_v=3N$ in this approximation. For small $\alpha T$, which is often the case
in the experimental temperature range, this implies $B\propto -T$, consistent with the experiments \cite{anderson}. We note
that experimentally, $B$ linearly decreases with $T$ at both constant volume and constant pressure \cite{anderson}. The
decrease of $B$ with $T$ at constant volume is due to the intrinsic anharmonicity related to the softening of interatomic
potential at large vibrational amplitudes; the decrease of $B$ at constant pressure has an additional contribution from
thermal expansion. Assuming $\omega^2\propto B_0+B_{ph}$ and combining it with $\left(\frac{\partial B_{ph}}{\partial
T}\right)_v=-\alpha(B_0+B_{ph})$ from above gives $\frac{1}{\omega}\frac{{\rm d}\omega}{{\rm d}T}=-\frac{\alpha}{2}$.
Putting this in Eq. (\ref{4d}) gives

\begin{equation}
C_v=3N(1+\alpha T)
\label{10}
\end{equation}

We see that the derived expression for $C_v$ is linear with $T$ and depends on $\alpha$ but not on $\omega$, unlike in Eq. (\ref{4d}). This result
follows from Eq. (\ref{4d}) as long as $\frac{{\rm d}B}{{\rm d}T}\propto B$ or $\frac{{\rm d}\omega}{{\rm
d}T}\propto\omega$.

From Eq. (\ref{10}), $\frac{C_v^l}{C_v^g}=\frac{1+\alpha_l T_g}{1+\alpha_g T_g}$ at $T_g$. Using it in Eq. (\ref{4b}) and
retaining only linear terms in $\alpha T$ (we find that the linearization and direct combination of Eqs. (\ref{4b}) and
(\ref{10}) give close values of $\frac{C_p^l}{C_p^g}$ below), we write

\begin{equation}
\frac{C_p^l}{C_p^g}=1+\gamma\alpha_g T_g(\epsilon-1)+T_g(\alpha_l-\alpha_g)
\label{13}
\end{equation}

Recalling that $\epsilon=\frac{(\epsilon_2+1)^2}{\epsilon_1+1}=\frac{\alpha_l^2}{\alpha_g^2}\frac{B_l}{B_g}$, we see that
Eq. (\ref{13}) relates $\frac{C_p^l}{C_p^g}$ to the changes of $\alpha$ and $B$ due to the presence of relaxational response
in the liquid and its absence in the glass.

Interestingly, Eq. (\ref{13}) predicts that temperature dependence of $C_p$ should follow that of $\alpha$. This is in agreement with recent simultaneous measurements of $C_p$ and $\alpha$ showing that both quantities closely follow each other across $T_g$ \cite{take}.

Importantly, the jump of heat capacity at $T_g$ in our theory takes place within the same single thermodynamic liquid phase, but below and above $T_g$ the liquid has different values of $\alpha$ and $B$ due to the freezing of LREs at $T_g$ where the liquid falls out of equilibrium. In this sense, our theory is purely dynamic. In contrast to previous glass transition theories \cite{dyre}, we do not discuss transitions between distinct thermodynamic phases, even though it may be tempting to invoke thermodynamic phase transitions, conventional or unconventional, in order to explain the heat capacity jump.

In Table 1, we show $\frac{C_p^l}{C_p^g}$ for several common glass-formers with both small and large $\frac{C_p^l}{C_p^g}$
in the range 1.1--1.8 \cite{angell}. Using the experimental values of $\gamma$, $T_g$, $\alpha_g$, $\alpha_l$, $B_g$ and $B_l$ \cite{daw,bom,bla,chr,moyn,rol,naoki,chang,taka,wu,carini,ding,whit}, we calculate $\frac{C_p^l}{C_p^g}$ using Eq. (\ref{13}). Given the approximations used, including $T_l=T_g$ and $V_l=V_g$ that underestimate $\frac{C_p^l}{C_p^g}$, Table 1 shows a reasonable agreement between the calculated and experimental values. The worse agreement for CKN is probably due to the fact that it is a solution \cite{angell}, for which our approximations are expected to be less successful. We further remark that the agreement is subject to uncertainties of $\gamma$, $\alpha$, $B$ and $\frac{C_p^l}{C_p^g}$ \cite{anderson,moyn} which are, moreover, taken from different experiments. For these reasons, we view Table 1 as an illustration that the differences between the existing values of $\alpha_g$ and $\alpha_l$ and between $B_g$ and $B_l$ are large enough to give the right magnitude of experimental $\frac{C_p^l}{C_p^g}$.

\begin{table*}[ht]
\begin{tabular}{ l l l l l l l l l l l l}
\hline
           & $\gamma$        & $T_g$          & $\alpha_l\cdot 10^4$ &
$\alpha_g\cdot 10^4$   & $B_l$          &       $B_g$   & $\frac{C_p^l}{C_p^g}$
& $\frac{C_p^l}{C_p^g}$ \\
           &                 & [K]            &[ K$^{-1}$]
&[K$^{-1}$]              & [GPa]          &       [GPa]   &   calc.
& exp.\\
\hline Glycerol   & 2.2 \cite{daw}             & 190 \cite{bom} & 5 \cite{bla} & 1 \cite{bla}     & 5.5 \cite{chr} & 9.9
\cite{chr}& 1.6                   &
1.8 \cite{angell} \\
PVAC       &0.6 \cite{moyn}  & 304 \cite{moyn} & 7.1 \cite{moyn} & 2.8 \cite{moyn}     & 2 \cite{moyn} & 3.5 \cite{moyn}&
1.3                   &
1.4 \cite{moyn} \\
OTP        & 1.2 \cite{rol}                 & 241 \cite{bom}        & 7.2 \cite{naoki}  & 3 \cite{naoki}     & 2.1
\cite{naoki}      & 3.7 \cite{naoki}
& 1.3                 & 1.5 \cite{angell,chang}  \\
OTP-OPP    & 1.3 \cite{rol}                 & 235 \cite{taka}       & 8.5 \cite{taka}   & 2.5\cite{taka}     & 2.9
\cite{taka}       & 5 \cite{taka} &
1.6                   &     1.5 \cite{taka}  \\
PS         & 0.5 \cite{wu,taka}   & 355 \cite{taka} & 6 \cite{taka} & 2.5 \cite{taka}      & 1.5 \cite{taka} & 2 \cite{taka}
& 1.2
&     1.3 \cite{taka}   \\
CKN        &0.9 \cite{moyn}  & 340 \cite{moyn} & 3.6 \cite{moyn} & 1.2 \cite{moyn}     & 7.6 \cite{moyn} & 15.9 \cite{moyn}&
1.2                   &
1.6 \cite{moyn} \\
B$_2$O$_3$ &0.3 \cite{moyn,carini}  & 550 \cite{moyn} & 4 \cite{moyn} & 0.5 \cite{moyn}     & 2.6 \cite{moyn} & 10
\cite{moyn}& 1.3                   &
1.4 \cite{moyn} \\
NaAlSi$_3$O$_8$ & 0.35 \cite{ding}          & 1100 \cite{whit}       & 0.54 \cite{ding}  & 0.23 \cite{ding}   & 20
\cite{ding}      & 40 \cite{ding}
& 1.05                   &     1.11  \cite{ding}\\
GeO$_2$    & 0.27 \cite{ding}               & 580 \cite{ding}       & 0.76 \cite{ding}  & 0.27 \cite{ding}   & 8.08
\cite{ding}      & 23.87 \cite{ding}
& 1.04                   &     1.08  \cite{ding,angell}\\
\\
\hline \label{table}
\end{tabular}
\caption{Experimental and calculated values of $\frac{C_p^l}{C_p^g}$. When more than one value of $\alpha$, $B$, $C_p$ or
$\gamma$ was available, we have taken the average. PVAC, PS, OTP, OPP and CKN stand for polyvinylacetate, polystyrene,
orthoterphenyl, orthoterphenyl phenol and Ca(NO$_3$)$_2$-KNO$_3$ mixture, respectively. The ratio of the last term in Eq.
(\ref{13}), $T_g(\alpha_l-\alpha_g)$, to the second term, $\gamma\alpha_g T_g(\epsilon-1)$, varies from 0.1 in glycerol and
0.3 in OTP-OPP to 2 in NaAlSi$_3$O$_8$ and 4 in GeO$_2$.}
\end{table*}

\section{Time-dependent effects}

The jump of heat capacity at $T_g$ in Eq. (\ref{13}) is due to different $\alpha$ and $B$ below and above $T_g$ due to the freezing of LREs at $T_g$. This reflects the empirical definition of $T_g$ as the temperature at which $\tau$ exceeds the observation time $t$, as in a glass transition experiment. We now remove the empirical constraint $\tau>t$, and consider the general case of arbitrary relationship between $\tau$ and $t$. This gives time-dependent properties of $\frac{C_p^l}{C_p^g}$.

Lets now consider two liquids at two different temperatures $T_1$ and $T_2$ with relaxation times $\tau_1$ and $\tau_2$ such that $T_2<T_1$ and $\tau_1<\tau_2$, and calculate the ratio of their heat capacities, $\frac{C_{p,1}}{C_{p,2}}$. The response of both liquids to pressure now includes viscous relaxational component due to LREs. Hence, we write $P=B_1\frac{v_{el}+v_{r,1}}{V_0}$ and $P=B_2\frac{v_{el}+v_{r,2}}{V_0}$, where $B_1$, $B_2$ are the bulk moduli and $v_{r,1}$, $v_{r,2}$ are the relaxational responses of the two liquids, respectively. Combining the two expressions gives

\begin{equation}
\frac{B_1}{B_2}=\frac{1+\frac{v_{r,2}}{v_{el}}}{1+\frac{v_{r,1}}{v_{el}}}
\label{14}
\end{equation}

Similarly, considering temperature-induced response in the two liquids gives

\begin{equation}
\frac{\alpha_1}{\alpha_2}=\frac{1+\frac{v_{r,1}}{v_{el}}}{1+\frac{v_{r,2}}{v_{el}}}
\label{15}
\end{equation}

\noindent where $\alpha_1$ and $\alpha_2$ are thermal expansion coefficients of the two liquids.

When relaxational response in the low-temperature liquid is absent, $v_{r,2}=0$, Eqs. (\ref{14}-\ref{15}) become Eqs. (\ref{2}-\ref{3}).
Note that $v_r$ and $v_{el}$ in Eqs. (\ref{14}-\ref{15}) are not the same because they are due to different effects of pressure and temperature. This difference will be accounted for below.

We now recall that relaxation of liquids at low temperature follows slow stretched-exponential form:
$q=q_0\left(1-e^{-\left(\frac{t}{\tau}\right)^\beta}\right)$, where $q$ is a relaxing quantity, $q_0$ is its amplitude, $t$
is observation time and $\beta$ is a stretching exponent \cite{dyre,angell,phillips}. $\beta$ decreases from 1 at high
temperature to 0.5--0.8 at $T_g$ \cite{casa}. Recently, we proposed that slow relaxation in liquids is a result of elastic
interaction between LREs via high-frequency elastic waves that they induce \cite{jpcm}. The stretched-exponential relaxation
follows as a result of this interaction \cite{ser}.

Hence, $\frac{v_{r,1}}{v_{el}}=q_0\left(1-e^{-\left(\frac{t}{\tau_1}\right)^{\beta_1}}\right)$ and
$\frac{v_{r,2}}{v_{el}}=q_0\left(1-e^{-\left(\frac{t}{\tau_2}\right)^{\beta_2}}\right)$, where $\beta_1$ and $\beta_2$ are stretching
exponents in both liquids. Therefore, as follows from Eqs. ({\ref{13}-{\ref{15}), time dependence of $\frac{C_{p,1}}{C_{p,2}}$ is given by the following three equations:

\begin{equation}
\frac{C_{p,1}}{C_{p,2}}=1+\gamma\alpha_2 T_2\left(\frac{\alpha_1^2}{\alpha_2^2}\frac{B_1}{B_2}-1\right)+\alpha_2 T_2\left(\frac{\alpha_1}{\alpha_2}-1\right)
\label{5}
\end{equation}

\begin{equation}
\frac{B_1}{B_2}=\frac{1+\epsilon_1\left(1-e^{-\left(\frac{t}{\tau_2}\right)^{\beta_2}}\right)}{1+\epsilon_1\left(1-e^{-\left(\frac{t}{\tau_1}\right)^{\beta_1}}\right)}
\label{6}
\end{equation}

\begin{equation}
\frac{\alpha_1}{\alpha_2}=\frac{1+\epsilon_2\left(1-e^{-\left(\frac{t}{\tau_1}\right)^{\beta_1}}\right)}{1+\epsilon_2\left(1-e^{-\left(\frac{t}{\tau_2}\right)^{\beta_2}}\right)}
\label{7}
\end{equation}

\noindent where we introduced relaxational amplitudes $\epsilon_1$ and $\epsilon_2$ as in Eqs. (\ref{2}-\ref{3}).

There are three times in Eqs. (\ref{5}-\ref{7}): $t$, $\tau_1$ and $\tau_2$. This sets three distinct regimes: (1)
fast regime $t\ll\tau_1\ll\tau_2$; (2) intermediate {\it glass transition} regime $\tau_1\ll t\ll\tau_2$ and (3) slow regime
$\tau_1\ll\tau_2\ll t$. Each regime sets a different mechanism governing the relationship between the heat capacities of the two liquids.

Regimes (1) and (3) both give $\frac{\alpha_1}{\alpha_2}=1$ and $\frac{B_1}{B_2}=1$ in Eqs. (\ref{6},\ref{7}), and therefore
$C_{p,1}=C_{p,2}$ from Eq. (\ref{5}), albeit for different physical reasons as discussed below.

Regime (1) corresponds to non-equilibrium states of both liquids and to non-equilibrium values of their $C_p$. In this regime, the terms in brackets in Eq. (\ref{6}) and Eq. (\ref{7}) are close to 0 because too little time elapses for any relaxation to take place. Physically, zero relaxational response in both liquids directly follows from Frenkel's idea that at short times, the response of a liquid is the same as that of a solid, i.e. purely elastic \cite{frenkel}. In this case, $C_{p,1}=C_{p,2}$ follows from the absence of relaxation in both liquids.

Regime (3) corresponds to equilibrium states of both liquids and to equilibrium values of their $C_p$. In this regime, the terms in brackets in Eq. (\ref{6}) and Eq. (\ref{7}) are close to 1 due to relaxational response acquiring its maximal value as both liquids reach their equilibrium state, giving $C_{p,1}=C_{p,2}$. Importantly, our theory predicts no difference between $C_{p,1}$ and $C_{p,2}$ in equilibrium (except for the residual difference due to $T_1>T_2$ and $V_1>V_2$ in Eq. (\ref{1}) discussed earlier and the difference in $C_v$ in Eq. (\ref{10}) related to a finite
temperature interval where $C_{p,1}-C_{p,2}$ is measured).

Regime (2) corresponds to the laboratory glass transition when the first liquid is in equilibrium ($\tau_1\ll t$) but the
second liquid is not ($t\ll\tau_2$). If the second liquid is defined as the glass, the change of heat capacity between the
liquid and the glass follows: $1-e^{-\left(\frac{t}{\tau_1}\right)^{\beta_1}}$ is close to 1 but $1-e^{-\left(\frac{t}{\tau_2}\right)^{\beta_2}}$ is close to 0 in Eqs. (\ref{6}-\ref{7}), giving $B_1=\frac{B_2}{\epsilon_1+1}$ and $\alpha_1=\alpha_2(\epsilon_2+1)$ as in Eqs. (\ref{2}-\ref{3}) and therefore non-zero $C_{p,1}-C_{p,2}$ in Eq. (\ref{5}), as in Eq. (\ref{13}). This is the important result of our theory.

We note that, by definition, the time range of regimes (1)-(3) is determined by the relationship between $\tau$ and the observation time $t$. In the calorimetry experiments in particular, $C_p$ is measured on the time scale $t=10^2-10^3$ s. Depending on $\tau$, regimes (1)-(3) can be too fast or too slow for the experiment. For example, if $\tau_1,\tau_2<100$ s, regimes (1) and (2) are too fast the experiment. On the other hand, regime (3) can be too slow for the experiment, and can last for astronomical times and longer. Lets consider SiO$_2$ glass at room temperature $T_r$=300 K. The activation energy barrier $U$ can be assumed temperature-independent, because SiO$_2$ is a strong liquid. Then, from $\tau(T_g)=\tau_0\exp(U/T_g)$ and $\tau(T_r)=\tau_0\exp(U/T_r)$, $\tau(T_r)=\tau_0\left(\frac{\tau(T_g)}{\tau_0}\right)^{\frac{T_g}{T_r}}$. Taking $\tau_0=$0.1 ps, $T_g\approx 1500$ K and $\tau(T_g)=10^3$ s, $\tau(T_r)=10^{67}$ s, approximately the fourth power of the age of the Universe. For $t>\tau(T_r)$, SiO$_2$ at room temperature is an equilibrium liquid, and shows no jump of heat capacity on cooling from high temperature related to the freezing of LREs.

Further, our theory explains two widely observed and important effects of glass transition. First, a well-known effect is
that $T_g$, defined as the temperature at which the jump of heat capacity takes place in the calorimetry experiment,
logarithmically increases with the quench rate $q$ (see, e.g., Refs. \cite{logar1,logar2}). According to our discussion above, the jump of heat
capacity at $T_g$ takes place when the observation time $t$ crosses liquid relaxation time $\tau$. This implies that
because $q=\frac{\Delta T}{t}$, $\tau$ at which the jump of heat capacity takes place is $\tau(T_g)=\frac{\Delta T}{q}$,
where $\Delta T$ is the temperature interval of glass transformation range. Combining this with
$\tau(T_g)=\tau_0\exp(U/T_g)$ (here $U$ is approximately constant because $\tau$ is nearly Arrhenius around $T_g$
\cite{jpcm}), we find:

\begin{equation}
T_g=\frac{U}{\ln\frac{\Delta T}{\tau_0}-\ln q}
\label{71}
\end{equation}

According to Eq. (\ref{71}), $T_g$ increases with the logarithm of the quench rate $q$. In particular, this increase is predicted to be faster than linear with $\ln q$. This is consistent with the experimental results \cite{logar1}. We note that this theory predicts no divergence of $T_g$ because the maximal physically possible quench rate is set by the minimal internal time $\tau_0$, Debye vibration period, so that $\frac{\Delta T}{\tau_0}$ in Eq. (\ref{71}) is always larger than $q$.

Second, Eqs. (\ref{5}-\ref{7}) predict the correlation of $\frac{C_p^l}{C_p^g}$ with liquid fragility \cite{angell,wang}.
Lets apply Eqs. (\ref{5}-\ref{7}) to the glass transition regime where $t\ll\tau_2$. Then, $1-e^{-\left(\frac{t}{\tau_2}\right)^{\beta_2}}=0$ in Eqs. (\ref{6}-\ref{7}), corresponding to the absence of relaxational response in the glass. Next, lets consider that $\Delta T_g$ is the temperature interval in which $\frac{C_p^l}{C_p^g}$ is measured so that $\tau_1=\tau(T_g+\Delta T_g)$ and $\tau_2\approx\tau(T_g)$. Then, because $\beta$ is nearly constant near $T_g$ \cite{casa}, $f=1-e^{-\left(\frac{t}{\tau_1}\right)^{\beta_1}}$ in Eqs. (\ref{5}-\ref{7}) can be expanded around $T_g$ as
$f\approx\frac{{\rm d}f}{{\rm d}\tau}\frac{{\rm d}\tau}{{\rm d}T}\Delta T_g$, where the derivatives are taken around $T_g$. From the definition of fragility $m=\frac{{\rm d}\log\tau}{{\rm d}\frac{T_g}{T}}\vert_{T=T_g}$, $\frac{{\rm d}\tau}{{\rm d}T}\vert_{T=T_g}\approx-\frac{\tau(T_g)}{T_g}m$. Next, $\frac{{\rm d}f}{{\rm d}\tau}\vert_{\tau=\tau(T_g)}\approx-\frac{1}{\tau(T_g)}$, giving $f=\frac{\Delta T_g}{T_g}m$, and Eq. (\ref{5}) becomes

\begin{equation}
\frac{C_p^l}{C_p^g}=1+\gamma\alpha_g T_g\left(\frac{\left(\epsilon_2\frac{\Delta
T_g}{T_g}m+1\right)^2}{\epsilon_1\frac{\Delta T_g}{T_g}m+1}-1\right)+\alpha_g\Delta T_g\epsilon_2 m
\label{8}
\end{equation}

If $m\frac{\Delta T}{T_g}\gg 1$, Eq. (\ref{8}) simplifies to

\begin{equation}
\frac{C_p^l}{C_p^g}\approx 1+\alpha_g\Delta T_g\epsilon_2\left(\gamma\frac{\epsilon_2}{\epsilon_1}+1\right)m
\label{9}
\end{equation}

Eqs. (\ref{8},\ref{9}) predict that $\frac{C_p^l}{C_p^g}$ increases with liquid fragility, provided other parameters do not
significantly change. This is consistent with experimental data from a large set of liquids \cite{angell,wang}. For a wider
range of chemically and structurally different liquids, the correlation holds within distinct families \cite{roland,huang}.

\section{Summary}

In summary, we observe that when a transition takes place between two distinct phases, anomalies in the thermodynamic
functions are explained by the phase transition theory, a well-understood topic in physics \cite{landau}. If a distinct
second phase can not be identified as in glass transition, the apparent anomalies can be explained in a picture that
does not invoke phase transitions and thermodynamics, but where the system stops relaxing on the experimental time scale.
Our approach explains time-dependent effects of glass transition, including widely observed logarithmic increase of $T_g$ with the quench rate and the correlation of heat capacity jump with liquid fragility.

We are grateful to M. Moore for discussions and to EPSRC and RFBR for support.

\end{document}